\documentclass[a4paper]{article}
\usepackage{ICASSPStyle}
\usepackage{graphicx}
\usepackage{amsmath}
\usepackage{enumitem}
\usepackage{ tipa }
\usepackage{ url}

\usepackage{caption}
\usepackage{wrapfig}

\usepackage{xcolor}

\newcommand{\etal}{{et~al.}}

\newcommand{\fakesubsection}[1]{\vspace{5pt}\noindent\underline{#1}:}

\title{Nonverbal Sound Detection for Disordered Speech} 
\name{\begin{tabular}{c}Colin Lea \qquad Zifang Huang \qquad Dhruv Jain$^{*}$\thanks{$^*$ University of Washington (work done during an internship at Apple)} \qquad Lauren Tooley \qquad Zeinab Liaghat \\ \qquad Shrinath Thelapurath \qquad Leah Findlater \qquad Jeffrey P. Bigham\end{tabular}}
\address{Apple Inc.}

\begin{document}
\maketitle

\begin{abstract}
Voice assistants have become an essential tool for people with various disabilities because they enable complex phone- or tablet-based interactions without the need for fine-grained motor control, such as with touchscreens. 
However, these systems are not tuned for the unique characteristics of individuals with speech disorders, including many of those who have a motor-speech disorder, are deaf or hard of hearing, have a severe stutter, or are minimally verbal. 
We introduce an alternative voice-based input system which relies on sound event detection using fifteen nonverbal mouth sounds like “pop”, “click”, or “eh.” 
This system was designed to work regardless of ones' speech abilities and allows full access to existing technology. 
In this paper, we describe the design of a dataset, model considerations for real-world deployment, and efforts towards model personalization.
Our fully-supervised model achieves segment-level precision and recall of 88.6\% and 88.4\% on an internal dataset of 710 adults, while achieving 0.31 false positives per hour on aggressors such as speech. 
Five-shot personalization enables satisfactory performance in 84.5\% of cases where the generic model fails.

\end{abstract}


\section{Introduction}
\label{sec:intro}

Many individuals with severe motor- or motor-speech disorders have limited communication ability and rely on ubiquitous technologies like phones and computers~\cite{NewAccessTech}. 
For people with motor impairments (e.g., carpal tunnel), assistive technology including voice control and eye tracking can be important parts of their daily life. 
Despite progress on disordered speech recognition \cite{Green2021,Harvill2021, Kim2017, Rudzicz2011}, commercial voice assistants have yet to be tuned for people with speech differences, so
individuals with ALS, Muscular Dystrophy, Traumatic Brain Injury or other motor-speech disorders may rely on physical switch controls (e.g., buttons, sip \& puff sensors, or joysticks) to interact with technology. 
These solutions can take orders of magnitude longer to accomplish the same tasks compared to people without motor disorders and may not be amenable to use in situations when an individual is laying in bed, outside of their wheelchair, or not at their desk~\cite{kane2020sense}. 

We present a system for nonverbal, sound-based interactions that people with a wide range of speech disorders can use to interact with mobile technology. 
The input is raw audio and the output is a set of discrete events triggered when a user makes one of fifteen mouth sounds, such as ``pop'', ``click'', or the phoneme $/$\textipa{i}$/$, which can be used to perform actions like ``select item'' or  ``go back'' on a mobile device. 
While conceptually simple, challenges arise when enabling robustness across wide vocalization ranges, achieving low-latency, and mitigating false positives from speech or background noises. 

Prior work consists of early prototypes that are not robust to the needs of all-day consumer technology \cite{Igarashi01,Funk20,VocalJoystick,VoiceDraw,Parrot.py} or use sounds that do not suit all disabilities \cite{cai2021voice,Talon}. 
Harada \etal~\cite{VocalJoystick,VoiceDraw} developed an early system for people with motor-speech disorders which predicted vowels such as $/$\textipa{u}$/$ and $/$\textipa{i}$/$ and associated them with computer mouse motions. 
While valuable, speech or background noises (wheelchair sounds, music) could easily produce false positives. 
Recently, Cai \etal~\cite{cai2021voice} introduced a system that is robust to the everyday needs of people with ALS, which solely detects the sound /a/, and is used to trigger actions like ``call for help.''
They prevent false positives by also training with environmental sounds and by requiring a user to repeat the sound twice within ten seconds. 
While robust, the post-processor prevents real-time use cases.
Talon Voice~\cite{Talon} and Parrot.py~\cite{Parrot.py} are voice control libraries designed for tech savvy individuals with motor disabilities. 
Talon has two detectors (``pop'' and ``hiss''), which can trigger system events on a computer, but which are not sounds users with certain oral-motor function can vocalize.
Parrot.py enables users to train custom detectors, but we find it is not robust to background sounds and can require tens or hundreds of training examples per detector.


We introduce a system that combines all of the benefits of the above work by:
(1) using a universal sound set (i.e., all speaking individuals should be able to trigger at least one sounds, regardless of speech, accents, or other vocal characteristics), 
(2) providing robustness for all-day usage (i.e., not falsely triggering when someone is talking, music is playing, or loud environmental sounds are occurring), and 
(3) having low-latency (i.e., system interaction is on-par with touch-based systems).
We describe a dataset, model, and a training scheme to improve robustness across variations in vocalizations, -- including via personalization -- and evaluate on data from individuals with and without motor-speech disorders.

\begin{table}[t]
    \small	
\begin{tabular}{|p{2.3cm}|p{5.9cm}| }
    \hline
    \textbf{Nonspeech} & \textbf{Definition}  \\
    \hline
    Click &  Tongue to roof of mouth (front), snap down. \\
    Cluck &  Tongue to roof of mouth (back), snap down. \\
    Pop &  Close lips tightly and release with quick blow. \\    
    \hline
    \textbf{Voiceless} & \textbf{Definition} \\
    \hline
    $/$\textipa{p}$/$: \underline{P}itch &  Close lips loosely and blow out. \\ 
    $/$\textipa{k}$/$:  \underline{K}ite & Touch back of tongue to roof of mouth, exhale. \\ 
    $/$\textipa{t}$/$:  \underline{T}eeth &  Open lips, teeth closed. \\     
    $/$\textesh$/$: \underline{Sh}oe (``sh'') & Push lips out with your teeth open and blow air. \\ 
    $/$\textipa{s}$/$:  \underline{S}nake & Open lips with your teeth closed and blow air. \\     
    \hline
    \textbf{Voiced} & \textbf{Definition}\\
    \hline
    $/$\textepsilon$/$: \underline{E}ffort (``eh'') & Open mouth, tongue raised, and start voicing. \\ 
    $/$\textschwa$/$: \underline{U}mp (``uh'')& Open mouth, teeth open, tongue slightly raised. \\     
    $/$\textipa{u}$/$: B\underline{oo} (``oo'') & Form an O with your lips. \\     
	$/$\textipa{m}$/$: \underline{M}om & Start voicing with lips closed. \\         
    $/$\textipa{i}$/$: \underline{Ea}gle (``ee'') & Open mouth with lips wide, tongue slightly raised, and start voicing. \\ 
    \hline
    \textbf{Diphones} & \textbf{Definition} \\
    \hline
    $/$\textipa{la}$/$: \underline{La}w & Touch tongue to roof of mouth. Start voicing and open mouth. \\ 
    $/$\textipa{m}\textschwa$/$:\underline{Mu}d(``muh'') & Close lips, start voicing, and open your mouth. \\ 
    \hline
\end{tabular}
    \caption{Nonverbal Sound List. IPA and English forms are used interchangeable in text for reader convenience.}
    \label{tab:sounds}
\end{table}



\section{Nonverbal Sounds}


\subsection{Sound Types \& Clinical Relevance}
With clinical guidance, we looked at prototypical examples of speech production for individuals with cerebral palsy, ALS, musclar dystrophy, multiple sclerosis, traumatic brain injury, and other conditions resulting in speech disorders. 
We identified 15 sounds spanning non-speech, voiced phones, unvoiced phones, and diphones as shown in Table~\ref{tab:sounds}.



Sounds were chosen based on features (i.e., voicing, nasality/resonance) that people with specific diagnoses are more likely able to produce while maintaining diverse locations of production in the oral cavity (palatal, alveolar, bilabial, velar) to ensure success for a large distribution of people. Vowels ``eh'', ``ee'' and ``oo'' were chosen for their spectral differences and because some (i.e., ``ee'' and ``eh'') may be more intelligible in individuals with ALS than other vowel choices \cite{Lee19}. The central vowel ``uh'' may be more easily produced for individuals with cerebral palsy and others with dysarthria  \cite{AnselKent92}.  ``Muh'' was chosen as a consonant-vowel (CV) production that is easier for individuals who tense their oral structures when initiating speech, while maintaining the central vowel ``uh''.  /m/ in isolation and in ``muh''  may be more clearly produced for people who may have hypernasality (i.e. due to flaccid dysarthria or when wearing BiPap for respiration).  Individuals who are unable to phonate consistently (e.g., due to respiratory incoordination, ventilator dependence, or a voice disorder) may be able to produce non-voiced phones such as /k/, /p/, /t/, /s/, ``sh'', or non-speech sounds ``click'', ``cluck'', and ``pop''.  

\begin{figure} 
	\centering
	\includegraphics[width=0.4 \paperwidth]{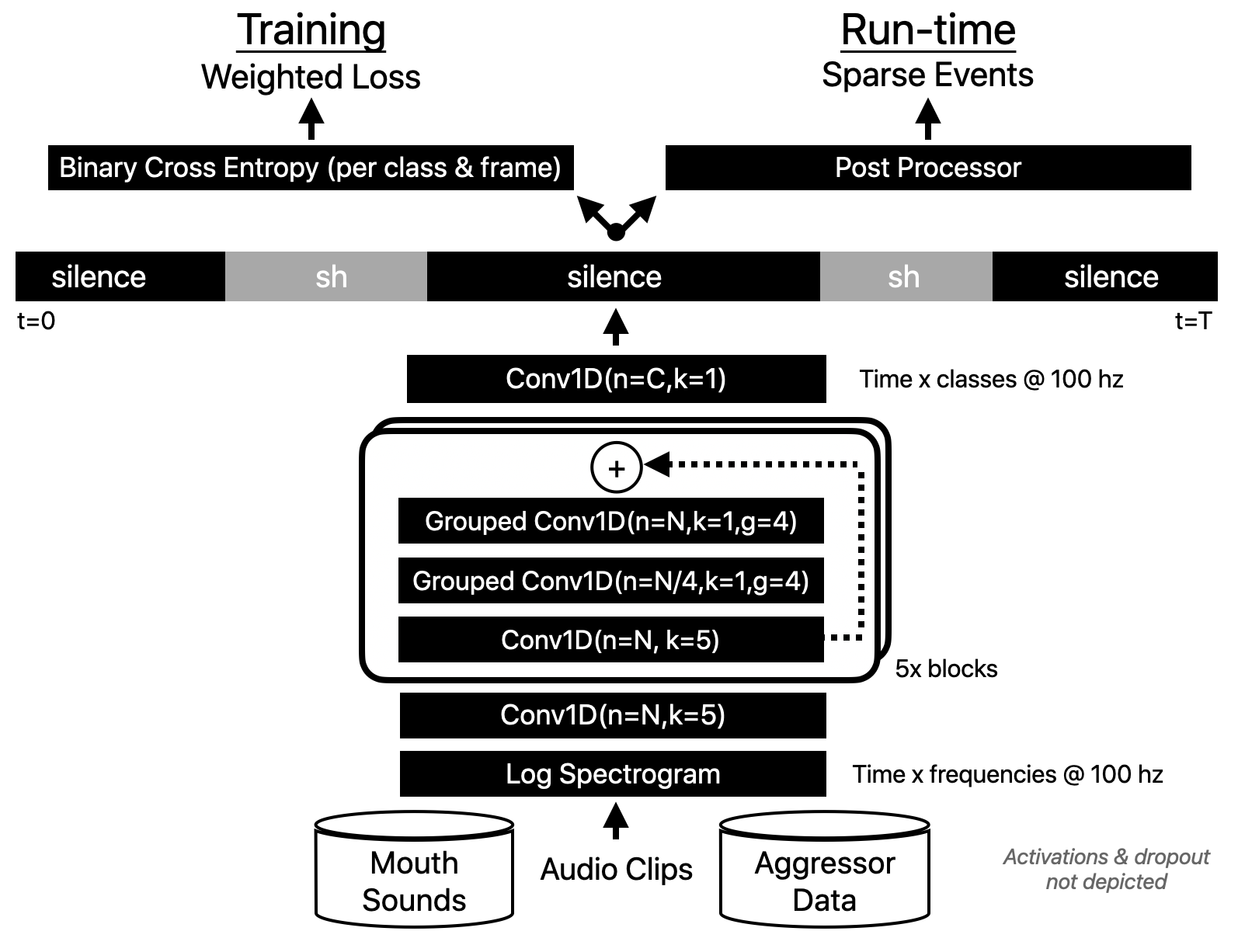}
	\caption{ During training, predict the probability of each sound per-frame, using mouth sounds and aggressor audio (speech, environmental sounds). At test time, take these probabilities and generate sparse events. 
	$k$=width, $g$=groups, $n$=nodes. }
	\label{fig:training}
\end{figure}

\subsection{Datasets}
\label{sec:data}

Despite the short duration of our sounds, there is large variation in pronunciation across accents, ages, genders, and vocal abilities. 
In contrast with (e.g., \cite{cai2021voice}) which train and evaluate isolated vowel detectors on public English speech datasets, we collected over 100k instances of isolated sounds using the protocol described below. 
We also use a large and diverse set of aggressor data, in the form of speech and environmental sounds, to prevent false detections in everyday situations.

\fakesubsection{Mouth Sounds} We collected audio from 710 non-disabled people with at least 40 participants each across demographics spanning accent/locale, age ranges (18+), gender (male, female, non-binary), device type (phone, tablet, wired or bluetooth headphones), and background environment (indoor or outdoor). 
Each person recorded audio clips of themselves repeating each sound type at least 10 times in a row with about one second of silence between vocalizations. 
Recordings were done at a ``close'' and ``far'' distance to simulate holding a device in hand and speaking into a tablet potentially mounted on a wheelchair or table. 

\begin{figure} 
	\centering
	\includegraphics[width=0.4 \paperwidth]{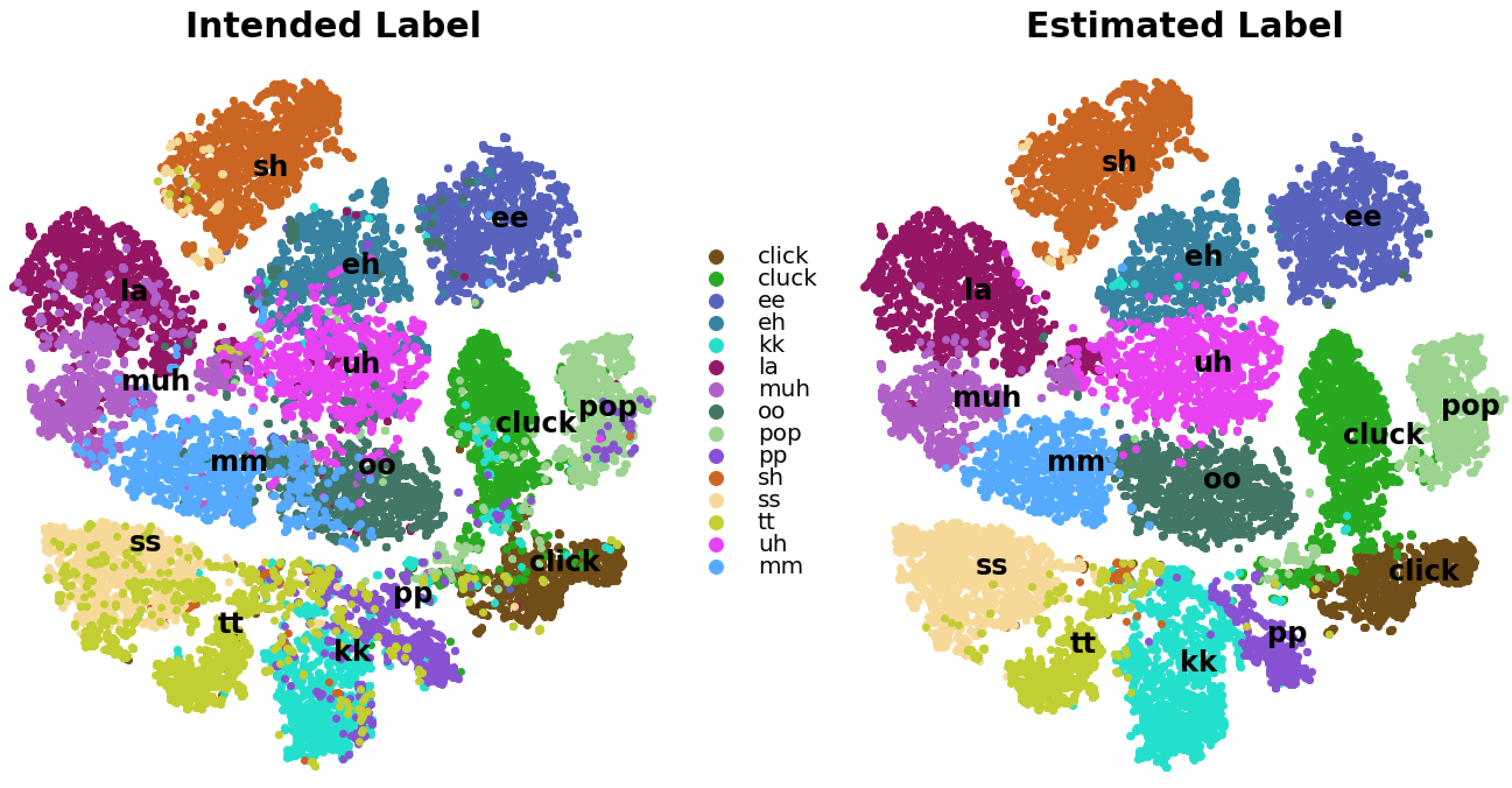}
	\caption{
	T-SNE visualization of sound event embeddings using the model in Section~\ref{sec:approach}. 
	Embeddings are colored using (Left) the sound type assigned by a participant and (Right) the sound type assigned by our model. 
	This is used to identify discrepancies in how people vocalized each sound and the label type.
    }
	\label{fig:pseudo_labels}
\end{figure}

Obtaining data across accents and physical locations is important. 
Early models trained on predominantly US accents achieved 24.7\% worse F1 score compared to the same models trained on people with nine accents (British, Chinese, French, German, Indian, Italian, Japanese, Spanish, US). 
One mode of variation was from people whose native language (e.g., Italian) only had five scripted vowels instead of the seven used in English (i.e.,  ``uh'' and ``eh'' are used interchangeably). 
We also found larger variation in how people from different countries tended to say each sound, regardless of our written and visual descriptions
(i.e., ``uh'' was sometimes pronounced ``oo'').
These discrepancies were apparent when listening to clips and when visualizing similarity of their sound embeddings, as 
shown via the T-SNE plot~\cite{tsne} in Figure~\ref{fig:pseudo_labels}. 
We automatically detected discrepancies using two rounds of Pseudo Labeling~\cite{lee2013pseudo} and found 9.8\% of self-described labels to be different than our prediction including 11.3\% from ``eh'' to ``uh'', 8.5\% ``uh'' to ``oo'', and 6.5\% ``ee'' to ``eh''. 
These clips were removed when training final models.


\fakesubsection{Aggressors} We train and evaluate using speech and non-speech datasets to mitigate false positives. For speech, we  use subsets of LibriSpeech which contains read speech~\cite{librispeech} (\textit{train-clean-100} and \textit{test-clean}), public podcast recordings of people with US and British accents, and 10 phrases from each participant in our mouth sounds collection. For non-speech data we rely on environment sounds from AudioSet~\cite{AudioSet} and internally collected recordings such as appliance sounds. Training clips are randomly sampled from each dataset, totaling 30 hours of speech and 20 hours of background sounds. 

\fakesubsection{Annotations}
Each mouth sound recording contains repeated instances of one sound type with silence in between.  
Frame-wise labels were generated by computing the energy in the audio signal and finding segments with minimum duration of 30 ms and whose relative energy exceeded one standard deviation from the mean. 
All frames within a given segment were labeled with the user-annotated sound type and all others were considered ``silence.'' 
Labels for speech clips were generated using a speech activity detector and all aggressor clip frames were labeled with the background class. 

\section{Low-latency Sound Detector}
\label{sec:approach}

Our system is visualized in Figure~\ref{fig:training}.
A preprocessor computes log spectrograms, a temporal convolutional network computes the probability that each frame contains a sound, and a post-processor takes probabilities and outputs sparse detections. 
At run time all modules are applied at 100 hz. 

\fakesubsection{Model Architecture}
Our model is a simple Temporal Convolutional Network, most similar to QuartzNet~\cite{quartznet}. 
The input 64 dimensional log mel-spectrograms generated from 16k hz audio with a 25 ms window and stride of 10 ms, resulting in a 100 hz sampling rate.
The first layers apply 1D convolutions (kernel size \texttt{k=5}) with \texttt{N=256} nodes. There are then five blocks of grouped (\texttt{g}) convolutions with the following pattern: \texttt{Conv1D(n=N,k=5,g=4)}, \texttt{LeakyReLU}, and a residual bottleneck consisting of \texttt{Conv1D(n=N/4,k=1)}, \texttt{LeakyReLU}, \texttt{Conv1D(n=N,k=1)}. 
Dropout is used after each activation. 
The network head consists of a \texttt{Conv1D(n=C, k=1)} with \texttt{Sigmoid} activation. 
Each frame's output is a vector of size $C=17$: 15 nonverbal sounds, a background class, and a speech class.
The receptive field is 270 ms.



\fakesubsection{Post-processing}
Many of our sounds are similar to what appear in everyday speech. 
We prevent false positives using a post-processor that aggregates background, speech, and nonverbal probabilities and outputs sparse events $(c,t)$ for class $c$ and time $t$. 
For each class, given probabilities $p_{c,t}$ for times $1...t$, generate an event if $p_{c}$ is greater than threshold $\theta_c$ for the most recent $\tau_c$ frames. 
No event is generated if the background or speech probabilities exceed $\theta_{bg}$ in the past 50 frames or if any class is detected within this time.

Post-processing parameters are optimized per-class to minimize the weighted F1 score on mouth sounds data, False Positive Rate on speech aggressor data, and latency. Optimal values range from $\theta_c \in [0.4, 0.6]$ and $\tau_c \in [7, 15]$ frames. 
Sounds including click and pop may be 50 ms whereas /u/ or /i/ may be 250 ms, and values of $\tau_c$ reflect this. 
If additional robustness is required, additional processing can be used to further reduce false positives by requiring silence after each sound, albeit at the cost of added latency. 






\subsection{Model Training}


Baseline models are trained using a binary cross entropy loss per-class.
Batches of 50\% mouth sound clips and 50\% aggressors are concatenated, with cumulative duration of $T$ frames, outputting $T$ log probability vectors, with a loss evaluated at 100 hz before the post-processing function. 
Boundaries of each segment are inflated by 50\% of the receptive field size (13 frames) to encourage the model to detect the onset and offset of a sound, where many of the constituent frames are ``silence''. 
This is equivalent to the temporal augmentation used by Meyer \etal ~\cite{fewShotHarvard}.


\fakesubsection{Personalization}
Our datasets contain predominantly non-disordered speech, and there is risk that the system does not work as well for users with severe speech differences. 
We investigated whether models could be personalized by fine-tuning on example vocalizations from a user.
We use $256$-dim embeddings from the pre-trained model above and fine tune on recordings of someone repeating the same sound one to five times. 
Weights in the final class-specific layer are updated using the automated labeling scheme described above and using a frame-wise binary cross entropy loss.
Models are trained using vocalizations from one recording and evaluated using a separate clip typically recorded 15 minutes later. 






\section{Experiments \& Analysis}
\label{sec:results}

\begin{figure} 
	\centering
	\includegraphics[width=0.415 \paperwidth]{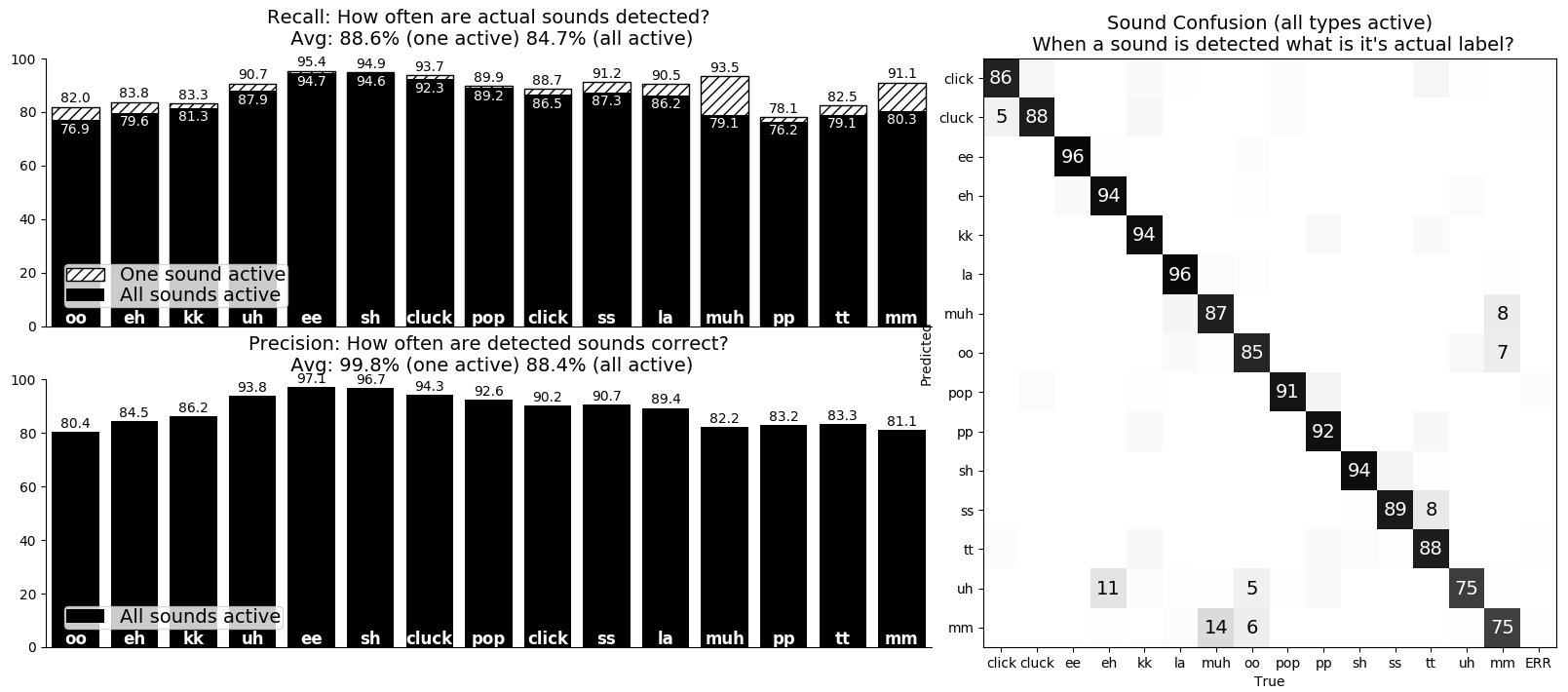}
	\caption{Segmental precision/recall on our 90 person non-disabled set. ``One active'' means only one sound type is enabled at a time. `All active'' means any class can be detected.
	}
	\label{fig:results}
\end{figure}

\fakesubsection{Baseline Results} Figure~\ref{fig:results} (top) shows segmental precision and recall metrics and a confusion matrix on a 90 person mouth sound evaluation set (5 male \& 5 female per accent). 
A segment is considered correct if the model detects the correct event anywhere between the start and end of a sound. 
In practice, someone using this type of feature may only use a few sound types per session; they likely will not need all 15 detectors at the same time. 
As such, results are shown for the extremes where only one detector is active (``one active'') and or all detectors (``all active''). 
The biggest discrepancy is for ``mm'' and ``muh'', which are often confused if both are active, but achieve high performance when used individually. 

\fakesubsection{Personalization} Experiments were performed on participants for whom the generic model fails (i.e., $F1<50\%$). Fine-tuning on one, three, or five examples from that user improves F1 by 55.8\%, 58.9\%, and  61.8\% on held out recordings. 84.5\% of sounds that could not be detected with the generic model could be detected after personalization with five samples. 
``click'' (74.3\%), ``pop'' (68.5\%), and ``oo'' (68.4\%) have the largest improvements.
Investigations with MAML and ProtoNets using \cite{learn2learn} did not yield significant improvements. 

\fakesubsection{Aggressors} Our final model, trained using ``positive'' mouth sounds and ``negative'' speech/background sounds, has 4.65 false positives (FPs) per hour 
on LibriSpeech test-clean. ``Sh'' and ``uh'' have higher rates (0.56 \& 0.74 FP/hour) whereas click and ``mm'' have zero. 
We trained the same model without these negative datasets and it has 303.9 FPs/hour.
Thus, this simple technique reduced the false positive rate by 98.4\% while losing only 0.6\% and 1.8\% precision and recall on mouth sounds data. 
A similar experiment on a 10.5 hour environmental sound set (e.g., kitchen noises) reduced the false positive rate from 238.5 to 0.225 FP/hour. Speakers and environments did not overlap in the training and test sets.


\fakesubsection{Latency}  The average system latency is 108$\pm$32 ms from the end of each vocalization to system detection. 
Extending the boundaries of each label as described in Section~\ref{sec:approach} improves sound onset detection and reduces latency by 33 ms.
Our model starts to detect sounds before they have been fully vocalized, which means that longer sounds such as /s/ or ``sh'' are sometimes detected before completion.
The computation time on an iPhone 12 is approximately 1 ms so the total amortized latency is within the range of typical touchscreen interactions (50-200 milliseconds~\cite{latency}). 



\fakesubsection{Motor-Speech Results} Recordings and feedback were collected from 28 people with speech differences resulting from cerebral palsy, muscular dystrophy, dysphonia, Parkinson's disease, or another motor-speech disorder.
Four had moderate-to-severe speech disorders as judged by speech intelligibility and the remaining had mild. 
Individuals tested a real-time version of this work and recorded themselves making each sound 10 times for quantitative evaluation. 
The average success rate (F1$\geq$50\%) was 82\%. Lowest performing sounds were $/k/$ (68\%), $/s/$ (72\%), $/t/$ (72\%). For 23 people, at least 10 of 15 sounds were successfully detected. 
Errors sometimes resulted when a sound was vocalized slowly relative to the non-disabled population or when an individual needed to vocalized sounds like $/t/$ as ``t-uh'' sometimes due to their speech difference.
For an individual with very low speech intelligibility only 3 of 15 sounds could be detected. 
For a user on a breathing apparatus, some sounds (e.g., ``sh'') did not work while the apparatus was active. This issue was mitigated by using other sounds that were not impacted such as $/k/$. 
Individuals with Parkinson's disease indicated that they would be interested in this feature during the times of the day when symptoms are most severe. 
Some individuals report needing to be close in proximity to the device due to a limited ability to vocalize loudly. 
We received positive feedback from people who have used our work in situations where they otherwise cannot interact with technology for mobility reasons (e.g., in bed or when not in their wheelchair).

\section{Conclusion}
\label{sec:conclusion}

We developed a system for nonverbal sound detection using triggers like pop and click that is robust to everyday interactions with background speech and environmental noise. This was designed to work for a wide range of vocal abilities and was validated on people with and without speech disorders. 



\bibliographystyle{IEEEbib}
\bibliography{strings,refs}

\end{document}